\documentclass[12pt]{article}
\usepackage{amsmath}
\usepackage[section]{placeins}
\usepackage{graphicx,hyperref}
\usepackage{dcolumn}
\usepackage{bm}
\usepackage{overcite}

\newcommand{\ket}[1]{|#1\rangle}
\newcommand{\bra}[1]{\langle#1|}

\newcommand{\eq}{\begin{equation}}
\newcommand{\fine}{\end{equation}}

\begin{document}

\title{Entanglement concentration after a multi-interactions channel}
\author{Eleonora Nagali$^1$, Fabio Sciarrino*$^{1,2}$, Francesco De Martini$^{1,3}$, \\
Miroslav Gavenda$^4$, Radim Filip$^4$\\
\small
$^1$Dipartimento di Fisica dell'Universit\'{a} ''La Sapienza'' and 
Consorzio Nazionale \\
\small
Interuniversitario per le Scienze Fisiche della Materia,\\
\small
Roma 00185, Italy.\\
\small
$^2$Centro di Studi e Ricerche ''Enrico Fermi'', Via Panisperna
89/A,\\
\small
Compendio del Viminale, Roma 00184, Italy\\
\small
*e-mail address: fabio.sciarrino@uniroma1.it\\ 
\small
$^3$Accademia Nazionale dei Lincei, via della Lungara 10, Roma 00165, Italy\\
\small
$^4$Department of Optics, Palack$\acute{y}$ University, 17. Listopadu 50,\\
\small
Olomuc 77200, Czech Republic} 
 
\maketitle

\begin{abstract}
Different procedures have been developed in order to recover entanglement
after propagation over a noisy channel. Besides a certain amount of
noise, entanglement is completely lost and the channel is called
entanglement breaking. Here we investigate both theoretically and experimentally an entanglement concentration protocol for a mixed three-qubit state outgoing from a strong linear coupling of two-qubit maximally entangled polarization state with another qubit in a completely mixed state. Thanks to such concentration procedure, the initial entanglement can be probabilistically recovered. Furthermore, we analyse the case of sequential linear couplings with many depolarized photons showing that thanks to the concentration a full recovering of entanglement is still possible.
\end{abstract}

\textbf{Keywords:} Entanglement, decoherence, quantum information protocols

\section{Introduction}
Quantum entanglement is a fundamental resource in quantum information, ensuring security of
key distribution and efficiency of quantum computing. Most of the quantum information protocols work best with maximally and pure entangled states, but unfortunately entanglement is extremely
sensitive to decoherence. Such difficulty imposes an extensive effort in order to recover a fairly good degree of entanglement after its interaction with noise, which is the main cause of the entanglement deterioration. In the last few years various schemes have been implemented in order to overcome this problem, especially in literature we refer to the \textit{distillation, purification} and \textit{concentration} schemes \cite{Benn96,Thew01,Durk02,Kwia01,Zhao03,Chen05,Vers04,Popp05,Pan03}. All these techniques deal with an entanglement recovery after a non-completely destructive interaction with noise in the communication channel. Even if a clear distinction between purification, distillation and concentration has not been precisely defined yet, it is possible to indicate as \textit{distillation} a procedure which increases the entanglement of a state. Instead the purification protocol increases the purity of the state, while the concentration protocol increases both the purity and the entanglement of a mixed state \cite{Thew01}.

In this work we present novel application of the three-photon entanglement concentration procedure in an entanglement-breaking process caused by a strong coupling with a noisy photon \cite{Sciar08}. This refers to a realistic situation, in which one photon from an entangled pair interacts trough a {\em coherent cross-talk} with a completely depolarized photon. Such photon acts as the environment which interacts with the signal trough a real noisy channel. The main novelty of such procedure is that the entanglement lost in the interaction is somewhat re-directed onto the signal photon pair by a 
measurement on the environmental photon \cite{Greg03,Busc05}. Usually, in most of the literature the environment, isolated with the system from the rest of the world, is considered as a pure state, which of course is easier to be completely controlled and characterized. We theoretically show that our protocol works even when the environment is in a mixed state. In this case the environment can been considered as a subsystem of a composite entangled state, giving rise to a wide variety of implementation in quantum information. 
\begin{figure}[h]
\centering
\includegraphics[scale=.35]{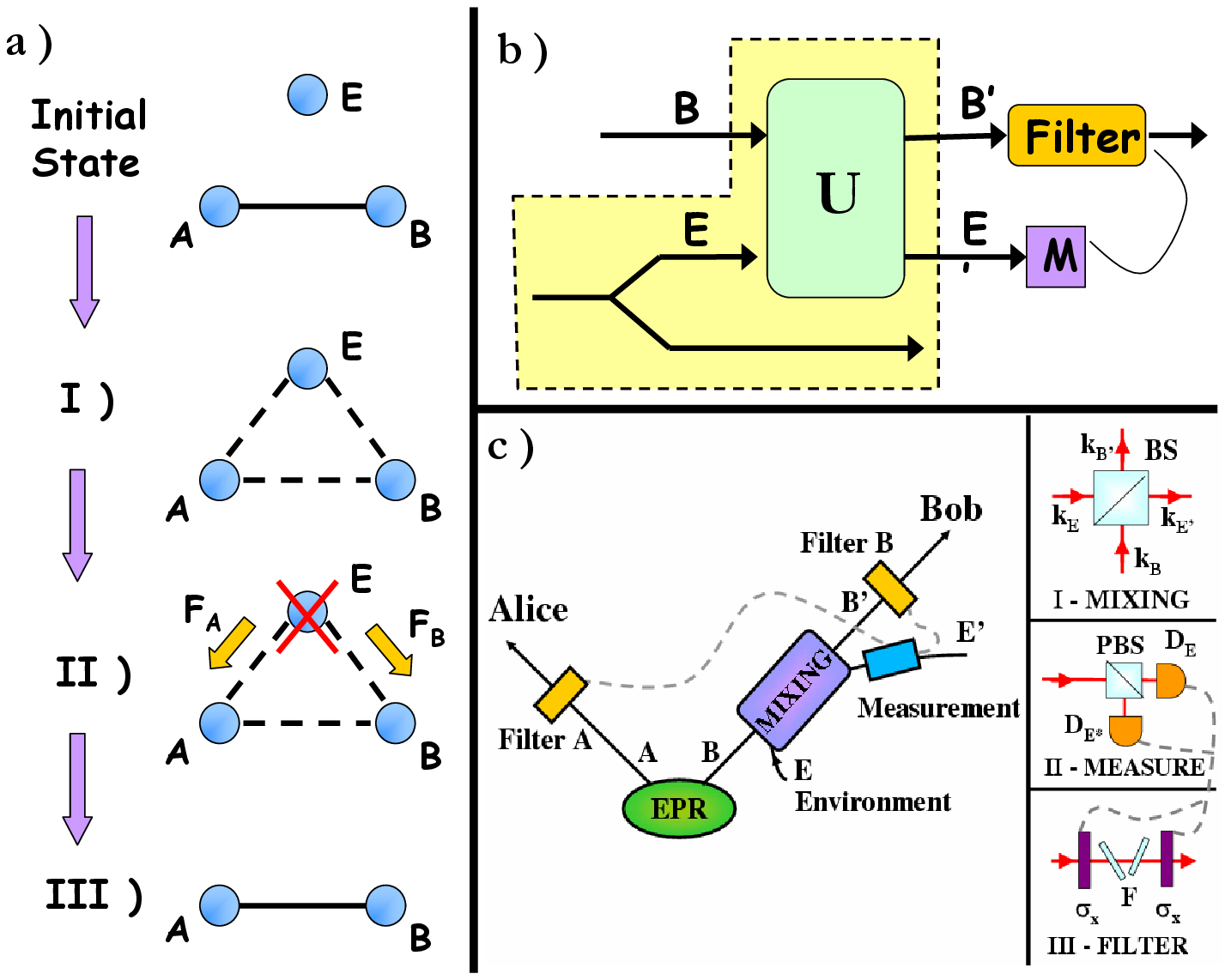}
\caption{Conceptual representation of the entanglement concentration protocol and its physical structure.\textbf{(a)}- The main steps of the concentration procedure are represented in order to enlighten the correlations established between the environment \textit{E}, Alice \textit{A}, and Bob \textit{B}.\textbf{(b)}- Generic coupling between the qubit B and the environment.\textbf{(C)}- Schematic representation of the concentration protocol. Generation by the EPR
source of the entangled photons shared between Alice
and Bob. On Bob mode is showed the interaction between the entangled photon
and the environment and the measure on the environmental photon that drives
the filtration process. On the right side are highlighted the experimental implementation of the three main steps of the protocol.}
\end{figure}
The experimental implementation of each single step of the concentration protocol has been carried out with the environmental photon in a fully mixed state which interacts with a maximally entangled state through a linear coupling. As further development of the entanglement concentration, we have carried an in-depth study on the interaction between noise and quantum information, generalizing the entanglement concentration process up to N coherent interactions between signal and noise.
As shown in \cite{Sciar08}, two different scenarios can be analyzed: whether the environmental photon and the signal one on the input mode of the BS are perfectly distinguishable or partially indistinguishable. 
Here we focus our attention only on almost \textit{indistinguishable} case, therefore we keep the traditional name of concentration. If the signal and the environmental photons are perfectly indistinguishable, the protocol exploits quantum interference
phenomena, analogously to the quantum state teleportation, and
asymptotically retrieves all the initial entanglement. On the other hand, even when the signal photon is in principle distinguishable from the environment but, due to technological limitations, can not be discriminated, it has been experimentally demonstrated that our protocol allows a partial recover of the initial entanglement.
\section{The Concentration protocol}
The overall dynamic of the protocol can be divided in three different steps:\\
\textbf{I) Coupling with environment.} Alice (A) and Bob (B) share a
polarization entangled state $|\Psi ^{-}\rangle _{AB}=\frac{1}{\sqrt{2}}(\ket{H}_{A}\ket{V}_{B}-i\ket{V}_{A}\ket{H}_{B})$, and the photon $B$
propagates over a noisy channel. The environment $E$ is
represented by a completely unpolarized photon described by the density
matrix ${\rho}_{E}=\frac{I}{2}$ that interacts with $B$ by a localizable linear coupling: a generic beam splitter BS with
transmittivity $T$ and reflectivity $R=1-T$. The present channel acts as a depolarizing channel, which means a balanced stochastic combination of the Pauli's operators $\sigma_{x}, \sigma_{y}, \sigma_{z}$.\\We point out that all the results that will be presented can be directly extended for general passive coupling between two modes \cite{Sciar08}. 
After the interaction on the beam splitter, three possible
situations can be observed: both the photons go to environment or to the
signal mode, or only a single photon is separately presented in signal and
environment. The first case corresponds simply to attenuation, while the
second case can be in principle distinguished by counting the number of
photons in the signal. Thus only the last case is interesting to be
analyzed. After the interaction process, the entanglement has been damaged
and, below a threshold value of $T$, no entanglement can be observed in the
output state.\\
\textbf{II)Measurement of the environment.} During the interaction on the BS, the initial entanglement between the pair photons is transferred to the three-photon mixed state entanglement between the environmental photon and the signal one. Such transfer is driven by the strength of the coupling, determined by the transmittivity T of the BS. Of course, due to the fact that we face to a mixed state environment, we refer to a correlation originated by the initial entanglement shared in the EPR pair, which is stored in the environmental photon after the interaction. Hence by properly measuring the environment it is possible to approach such correlation and restoring it into the entangled signal pair. \\The photon propagating on the environmental mode is measured after a
polarization analysis realized through a polarizing beam splitter (PBS).\\
\textbf{III)Filtration.} A interesting role in the protocol is represented by a
filtration performed both on Alice and Bob modes. Such operation leads to a higher entanglement of the bipartite system, at the cost of a lower probability of
implementation. We indicate the
attenuation over the mode $k_{i}$ for the $V$ polarization as $A_{i}$: $\ket{V}_{i}\rightarrow \sqrt{A_{i}}\ket{V}_{i}$. The complete protocol
implies a classical feed-forward on the polarization state of the photon
belonging to the Bob's mode depending on which detector on environmental mode fires. This
conditional operation could be realized adopting the electronic scheme
experimentally demonstrated in \cite{Giac02,Scia06}. In the present
experiment the active procedure has not been applied, hence
the efficiency of the overall procedure is reduced by a factor 2.\\
For each step of the protocol, we reconstruct the two qubit density matrix through
the quantum state tomography procedure \cite{Jame01}.
\section{Dynamics of the entanglement-breaking process}
After the coupling between entangled signal and the environmental photon on the beam-splitter, the
whole state is described by the three-qubit density matrix shown in Fig.(2-a). As underlined previously, and in-depth
 analysis of such matrix allows to achieve a better knowledge on the quantum
 correlation dynamics, shown in Fig.(2-b). The parameter $T$ establish the degree of correlation damaging between the photon A and the B one that is, how the noise corrupts the signal. A high value of
 transmittivity $T$ simulates the case of a weakly noisy channel, which does not
 destroy completely the initial information (red line in Fig.(2-b)). Usually this is the regime in which all the already implemented protocols work. Lower the value of transmittivity $T$, greater is the correlation established between the environment and the photon A, at cost of a concurrence reduction in the initial entangled state described by $\rho_{AB}^{IN}=\ket{\Psi^-}_{AB}\bra{\Psi^-}$. When a balanced beam splitter is
used for the coupling process, the concurrence referred to the density matrix
of the environmental photon and the Bob's one achieves its maximum value, since it
is related to the generation of a maximal entangled state shared between B and E.
As a consequence the initial entanglement between Alice and Bob is completely lost and transferred to Bob and to the environmental photon. This transferred entanglement is exploited in the concentration scheme. The entanglement between $B$ and $E$ generated by the coupling is continuously converted to entanglement between $A$ and $E$. For the strong coupling $T<1-1/\sqrt{3}$, the $A-E$ entanglement is redirected back to $A-B$ entanglement. However for $T\in(1-1/\sqrt{3},1/\sqrt{3})$ there is no entanglement between $A-B$ and $A-E$, and the non-local correlation survives only locally between $B-E$. Such entanglement is crucial for the concentration procedure, since together with classical correlations helps to restore $A-B$ entanglement.
\begin{figure}[h]
\centering
\includegraphics[width=9cm]{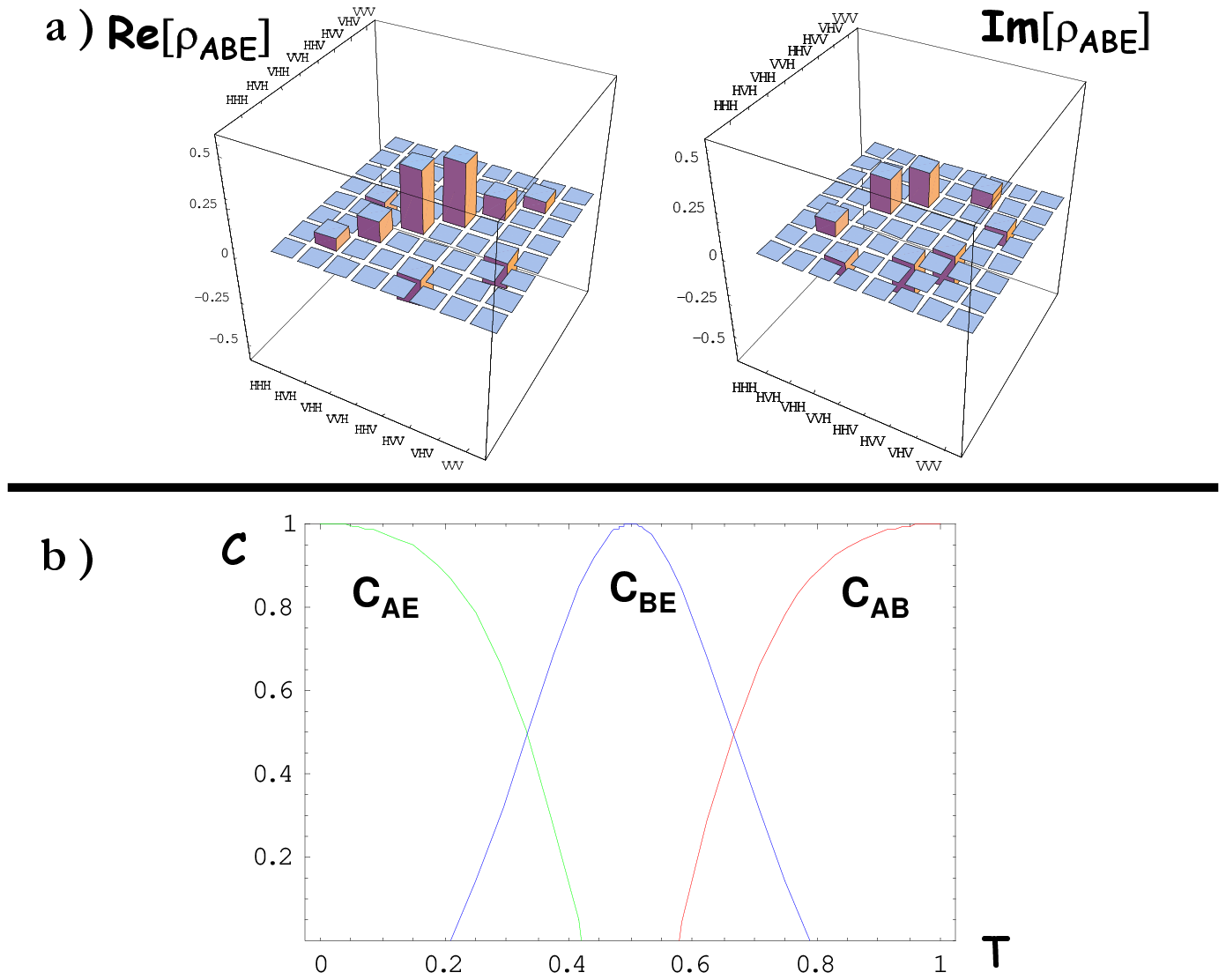}
\caption{\textbf{(a)}- Theoretical density matrix $\rho_{ABE}$ representing the three-qubit state after the coupling process on $BS$ for $T=0.3$. \textbf{(b)}- Plot of the concurrence C for different values of the transmittivity T of the BS. Each concurrence refers to the correlation established between two subsystems.}
\end{figure}
\section{Experimental implementation}
In order to exploit the regime where the concentration protocol works best, we
consider the signal photon and the noise one spatially, spectrally and temporal
perfectly indistinguishable in the coupling process.
A good spatial overlap is achieved by selecting the output modes through single mode fibers. The entangled state $\ket{\Psi ^{-}} _{AB}$, that is the signal to be transmitted through the noisy channel, has been generated by a NL crystal of $\beta$-barium borate (BBO) cut for type-II phase-matching.\\
\textbf{I)Mixing.} In the indistinguishable photons regime, we indicate with ${\sigma}%
^{I}_{AB}$ the density matrix after the mixing on the $BS$, which is separable
for $T< 1/\sqrt{3}$, as shown in Fig.3. The output state is found to be a Werner state, which is a mixture of the singlet state and the fully mixed one \cite{Wern89}. Without
 tracing over the environmental mode, the three photons state is described by the
 theoretical density matrix shown in Fig.(2-a).\\
\begin{figure}[h]
\centering
\includegraphics[scale=.28]{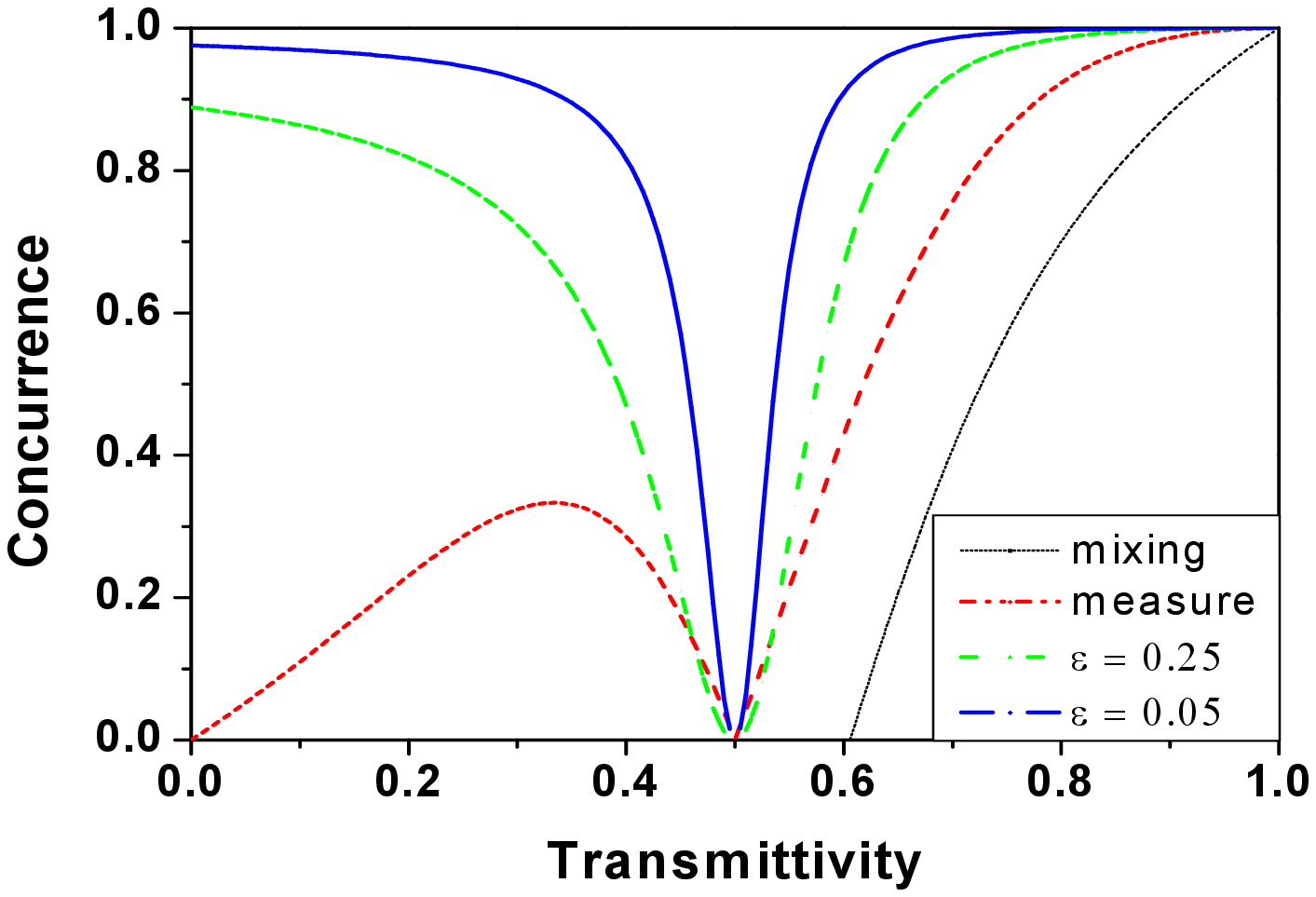}
\caption{Concurrence for indistinguishable
photons - without measurement (dotted line), with measurement (dashed
-dotted line), with measurement and LOCC filtration $\varepsilon =0.25$
(dashed line), with measurement and LOCC filtration $\varepsilon =0.25$
(continuous line).}
\end{figure}
\textbf{II)Measurement.} A measurement is carried out on the
environmental mode. When the result $|H\rangle_{E}$ is obtained, ${\sigma%
}^{I}_{AB}$ evolves into ${\sigma}^{II}_{AB}$:
\[{\sigma}^{II}_{AB}=\frac{1}{4P_{II}}
\left(
\begin{array}{cccc}
0 & 0 & 0 & 0 \\
0 & T^2 & -T(T-R) & 0 \\
0 & -T(T-R) & (T-R)^2 & 0 \\
0 & 0 & 0 & R^2
\end{array}
\right)
\]
where $P_{II}$ is the probability of implementation success of the process. The Werner state is hence conditionally transformed into a maximally entangled state (MEMS) \cite{Vers01}: see Fig.4. In particular, it is found $C_{II}=\frac{T|T-R|}{2P_{II}}$ and $P_{II}=\frac{T^2+(T-R)^2+R^2}{4}$.\\
Let us now compare the theoretical predictions to experimental results. The experimental generation of perfectely indistinguishable photons has to face some unavoidable technological limitations. In order to account this situation, for the experimental data analysis we have considered the state shared between Alice and Bob as a mixture between perfectely indistinguishable photons, weighted by the degree of indistinguishability $p$, and distinguishable photons. Such state leads to a reduction on the correlation restored by our protocol, due to a lower interference between signal and environment.\\
In order to estimate the parameter $p$, we have carried out an
Hong-Ou-Mandel interferometer experiment \cite{Hong87} adopting a BS with $T=0.5$, which has yielded to $p=(0.85\pm 0.05)$.\\
The experimental results are shown in Fig.(5-\textbf{II}) for $T=0.4$. The entanglement is restored with a
concurrence equal to $\widetilde{C}_{II}=(0.15\pm 0.03)>0$ to be compared with $C_{II}=0.22$; the fidelity with the theoretical state is $F=(0.96\pm 0.01).$\\
\textbf{III)Filtration} Due to the strong unbalancement of ${\sigma}^{II}_{AB}$ a filter
is introduced either on Alice or Bob mode, depending on the value of $T$: Fig.3. If $T>|2T-1|$, the filter acts on Bob mode, as $\ket{H}_{B}\rightarrow{\frac{2T-1}{T}\ket{H}_{B}}$, while if $T<|2T-1|$, the filter acts on Alice mode, as $\ket{V}_{A}\rightarrow{\frac{T}{2T-1}\ket{V}_{A}}$.\\
In order to properly apply the filtration, it is necessary to estimate the value of $T$. Hence the filtration procedure can be successfully applied after a characterization of the communication channel. Over a time period during which the transmission channel is a stationary one, is possible to quantify the "transmittivity" (amount of noise introduced), comparing the signal at the input and output of the channel.
\begin{figure}[h]
\centering
\includegraphics[scale=.35]{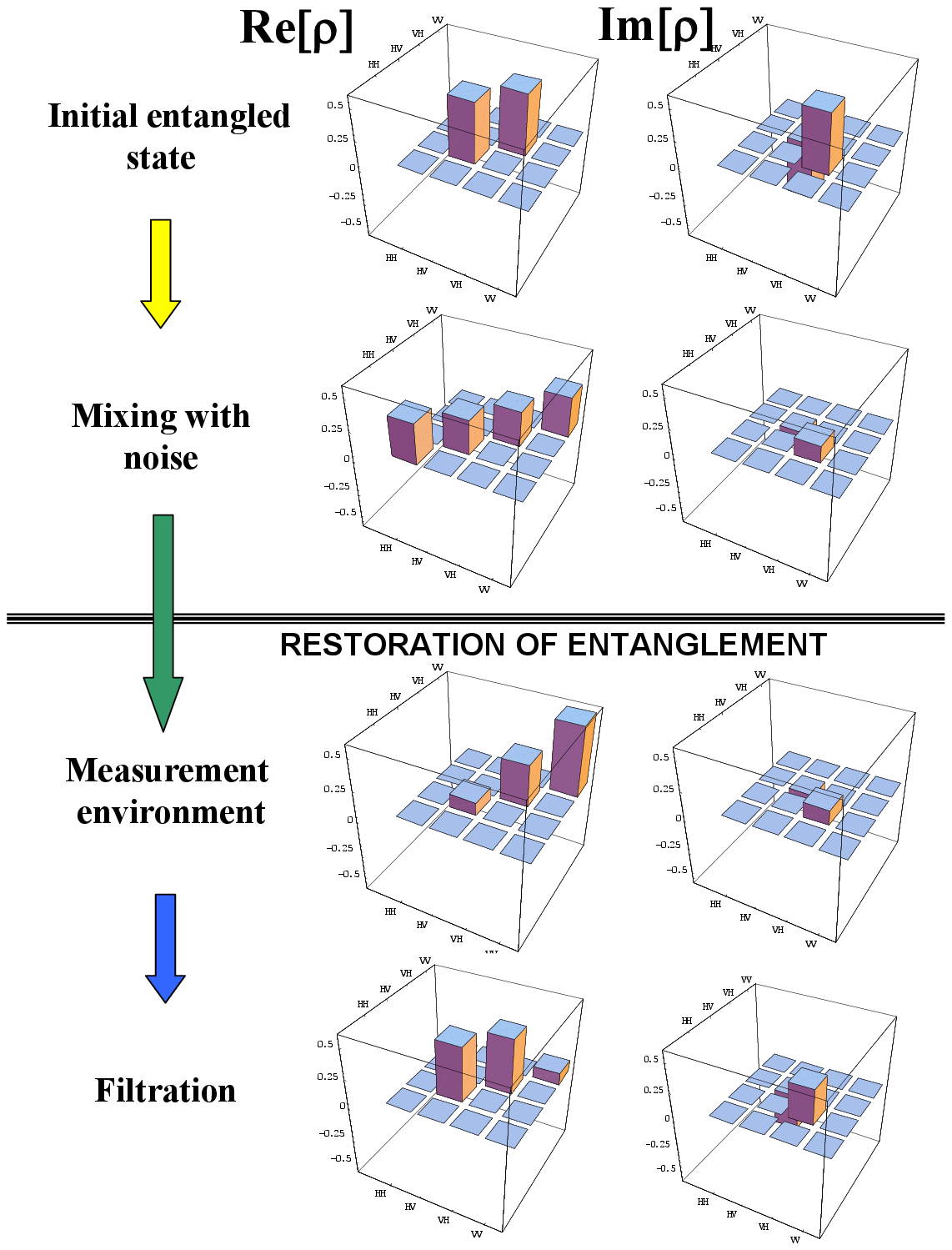}
\caption{Theoretical density matrices for perfectly indistinguishable photons interacting on the $BS$.}
\end{figure}
In order to increase the concurrence, a second filter is inserted on both Alice and Bob mode, which attenuates the vertical polarization component: $\ket{V}\rightarrow \sqrt{\varepsilon}\ket{V}$. The filters transform the density matrix into ${\sigma}^{III}_{AB}$:
\[{\sigma}^{III}_{AB}=\frac{1}{4P_{III}}
\left(
\begin{array}{cccc}
0 & 0 & 0 & 0 \\
0 & \varepsilon\alpha & -\varepsilon\alpha & 0 \\
0 & -\varepsilon\alpha & \varepsilon\alpha & 0 \\
0 & 0 & 0 & \varepsilon^{2}\delta
\end{array}
\right)
\]
where $\alpha=T^{2}$, $\delta=\left( \frac{TR}{R-T}\right) ^{2},$ for $T<|2T-1|$ and $\alpha=(2T-1)^{2}$, $\delta=R ^{2},$ for $T>|2T-1|$. The concurrence has value $C_{III}=\frac{2\varepsilon\alpha}{2\varepsilon\alpha+\varepsilon^{2}\delta}$ while the probability reads $P_{III}=\frac{2\varepsilon\alpha+\varepsilon^{2}\delta}{4}$. In Fig.3 are reported the concurrence values for $\varepsilon=0.05$ and $\varepsilon =0.25$. Hence in the limit of asymptotic filtration ($\epsilon \rightarrow 0$), the concurrence reaches unity except for $T=1/2$. Since maximally
entangled state can be approached with an arbitrary
precision  \cite{Vers01b}, the entanglement breaking channel can
be transformed into a secure channel for key distribution.\\
\begin{figure}[t]
\centering
\includegraphics[scale=.25]{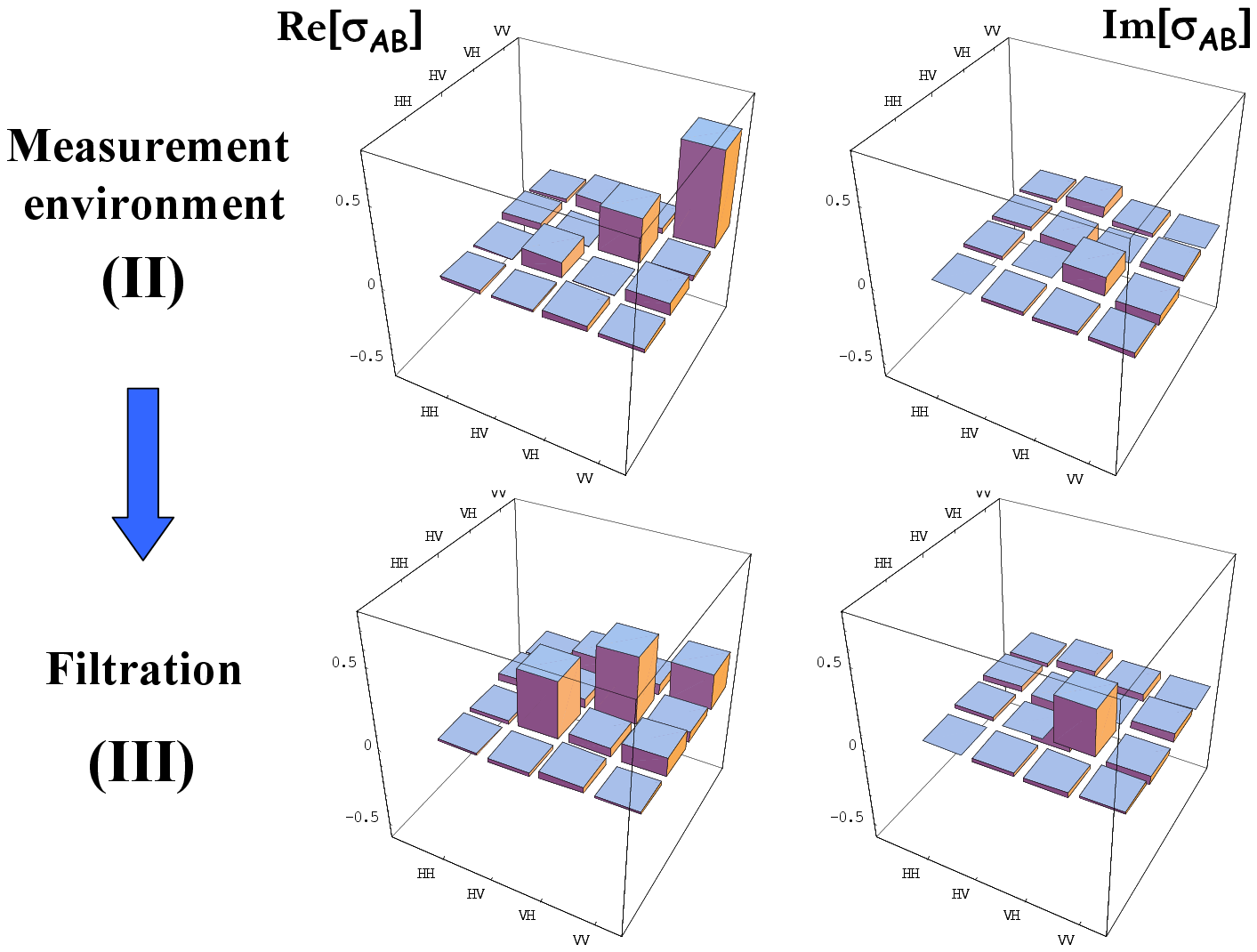}
\caption{Experimental density matrices for indistinguishable photons for different steps of the concentration protocol. Since a perfect indistinguishability between noise and signal is almost impossible to be achieved experimentally, we have estimate the degree $p$ of indistinguishability to be equal to $p=0.85$.}
\end{figure}
Applying experimentally the filtration with the parameters $A_{A}=0.12$, and $A_{B}=0.30$
we have obtained the state shown in Fig.(5-\textbf{III}). Hence we have measured a higher
concurrence $\widetilde{C}_{III}=(0.50\pm 0.10)>\widetilde{C}_{II}$ while the expected theoretical value is $C_{III}=0.47$. The filtered state has $F=(0.92\pm 0.04)$ with the theoretical expectation. This is a clear experimental demonstration of the entanglement concentration with the help of the local filtration in our realistic scenario.\\ 
\section{Multiple couplings with noisy photons}
Let us to generalize the entanglement concentration process up to N interactions between the signal and N environmental photons coherently coupled to the signal. We consider all couplings to be linear, that is an interaction on a beam splitter (see Fig.6). In particular, in order to generalize such calculation, we indicate with $BS_i$ the $i-th$ beam splitter with a transmittivity $T_i$ and reflectivity $R_i=1-T_i$. Furthermore, we consider $T_{i}\neq T_j$ for $i\neq j$. While it can be deduced that the concentration protocol can ideally work properly if all the three steps are applied for each interaction, our calculation refers to the case in which the filtration procedure is applied only after all the $N$ interactions between signal and noise. As a first step, we study the case of a double interaction between noise and signal and then we will consider $N$ beam splitters on the signal mode.\\
\begin{figure}[h]
\centering
\includegraphics[width=8.5cm]{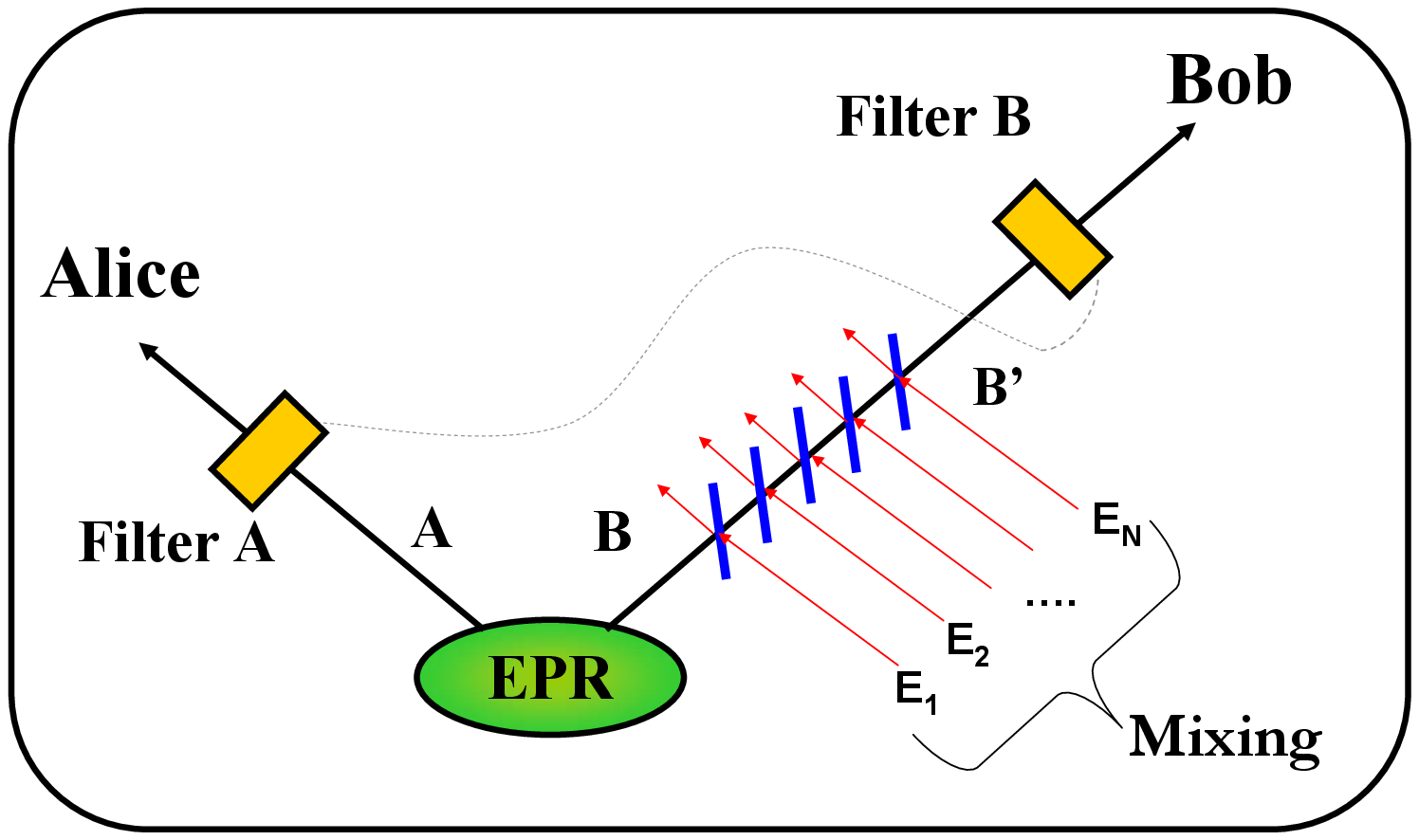}
\caption{Schematic representation of the entanglement concentration joined procedure implemented N times.}
\end{figure}
\textbf{I)Mixing.} After the first interaction between signal and the noise photon $E_1$, and the projective measurement $|H\rangle\langle H|$ on the environmental mode, the transmitted signal interacts with a second noise photon $E_2$ on a beam splitter characterized by transmittivity $T_{2}$.
From the previous calculation, after the interaction on $BS_{1}$ and projective measurement 
the state exhibits the concurrence $C^{(1)}=C_{II}$.
In the next step, it interacts with a second environmental photon described by the mixed state $\rho_{E2}=\frac{I}{2}$. Hence the overall state interacting on the beam splitter $BS_{2}$ reads
$\rho_{ABE2}=\rho_{AB_1}^{II}\otimes \rho_{E2}$. The three-qubit matrix ${\rho}_{ABE2}$ evolves in accordance with the transformation rules of the beam splitter, written for a generic transmittivity $T_i$ as:
\begin{itemize} 
\item $\ket{\phi}_{B_i}\ket{\phi}_{E}\rightarrow{(T_i-R_i)\ket{\phi}_{B_{i+1}}\ket{\phi}_{E}}$
\item $\ket{\phi}_{B_i}\ket{\phi^{\bot}}_{E}\rightarrow{T_i\ket{\phi}_{B_{i+1}}\ket{\phi^{\bot}}
_{E}-R_i}\ket{\phi^{\bot}}_{B_{i+1}}\ket{\phi}_{E}$
\end{itemize}
where $B_i$ and E indicates respectively the input modes for the signal and environmental photon,  while $B_0=B$ and $B_N=B'$.

\textbf{II)Measurement.} After two steps of interaction, the second environmental mode is projected on $|H\rangle$ state, hence the matrix $\rho_{ABE_2}^{IN}$ evolves to $\rho_{AB_2}^{II}$. 
Hence we get the matrix $\rho_{AB}^{M2}$:
\begin{equation}
{\rho}_{AB_2}^{II}=\frac{1}{8P^{(2)}_{II}} \left(
\begin{array}{cccc}
0 & 0 & 0 & 0\\
0 & a & -\sqrt{ab} & 0\\
0 & -\sqrt{ab} & b & 0\\
0 & 0 & 0 & c \\
\end{array}
\right )\\
\end{equation}
where $a = T_{1}^2 T_{2}^2$, $b = (T_{1}-R_{1})^2 (T_{2}-R_{2})^2$, $c = (T_{1}-R_{1})^2 R_{2}^2+R_{1}^2 T_{2}^2$ and $P^{(2)}_{II}=\frac{a+b+c}{8}$ is the normalization factor. It is possible to observe that $\rho_{AB_2}^{II}$ describes a MEMS state, where the concurrence can be expressed as:
\begin{equation*}
C^{(2)}_{II}=\frac{\sqrt{ab}}{4P^{(2)}_{II}}
\end{equation*}
Naturally, concurrence $C^{(2)}_{II}$ is vanishing for $T_{1}(T_{2})=0,\frac{1}{2}$ due to bunching of photons on the linear coupling. But in all the other cases, the entanglement was restored by the concentration measurement.
In the generalized case, where the signal interacts N-times with an environmental completely mixed photon, the matrix form allows a widespread expression of the density matrix which describes the quantum state after the N-th interaction. In particular, after the projection on $\ket{H}_{EN}$ state on each environmental mode, it is possible to demonstrate that the state on the output mode of the $N-th$ beam splitter is described by the matrix:
\begin{equation*}
{\rho}_{AB_N}^{II}=\frac{1}{2^{(N+1)}P^{(N)}_{II}} \left(
\begin{array}{cccc}
0 & 0 & 0 & 0\\
0 & A_{N} & -\sqrt{A_{N}B_{N}}& 0\\
0 & -\sqrt{A_{N}B_{N}} & B_{N} & 0\\
0 & 0 & 0 & C_{N} \\
\end{array}
\right )\\
\end{equation*}
where $P^{(N)}_{II}=\frac{A_{N}+B_{N}+C_{N}}{2^{(N+1)}}$ is the probability of success for the implementation of the $N-th$ measure. The parameters $A_{N}, B_{N}, C_{N}$ are function of $T_{i}$ and $R_i$ in the form:
\begin{itemize}
 \item $A_{N}=\prod_{i=1}^{N}T_{i}^2$
 \item $B_{N}=\prod_{i=1}^{N}(T_{i}-R_{i})^2$
 \item $C_{N}=R_{N}^2 B_{N-1}+T^2_{N} C_{N-1}$ with $C_{1}=R_{1}^2$
 \end{itemize}
The form of $\rho_{AB_N}^{II}$ shows that the matrix describing the state after the N-th interaction and measurement is, analogously to the matrix $\sigma_{AB}^{II}$, a MEMS state. 
Hence the concurrence after N interactions $C=C^{N}_{II}$ reads:
\begin{equation*}
C^{(N)}_{II}=\frac{\sqrt{A_{N}B_{N}}}{2^NP^{(N)}}
\end{equation*}
In Fig.7, the concurrence is reported as a function of $N$. As expected, as $N$ increases, it corresponds to a drastic reduction of the concurrence, if a proper filtration process is not applied. \\
\begin{figure}[h]
\centering
\includegraphics[width=9cm]{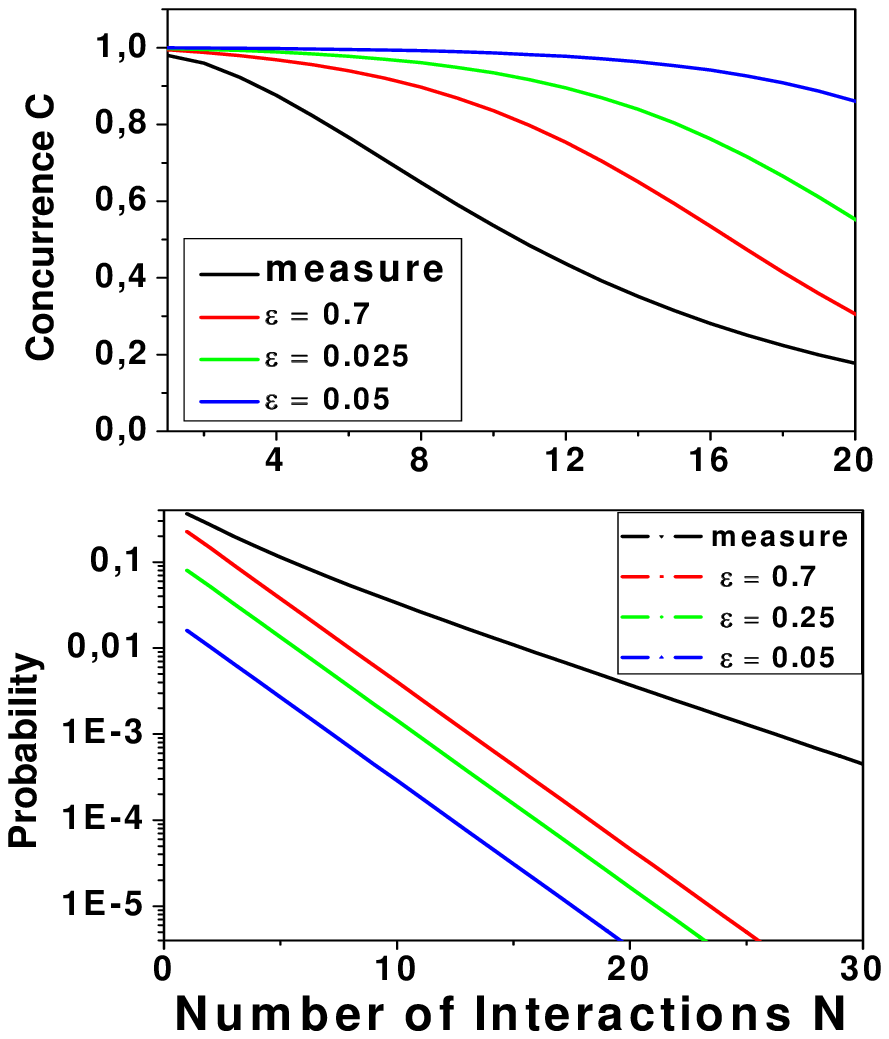}
\caption{Concurrence and probability of implementation for different interactions (N) for $T=0.1$. The black line indicates the probability after the N-th measurement (M) on the environmental mode, while the other lines represent the probability after the filtration process for different values of $\varepsilon$.}
\end{figure}
\textbf{III) Filtering.} In order to achieve a concurrence equal to one, it necessary to introduce a filtration which acts both on Alice and Bob mode as:
\begin{equation*}
\ket{V}_{A,B}\rightarrow \sqrt{\varepsilon}\ket{V}_{A,B}
\end{equation*}
\begin{equation*}
\ket{H}_{A}\rightarrow \sqrt{\frac{B_{N}}{A_{N}}}\ket{H}_{A}
\end{equation*}
which gives rise to a balanced density matrix:
\begin{equation*}
{\rho}_{AB_N}^{III}=\frac{1}{2^{(N+1)}P^{(N)}_{III}} \left(
\begin{array}{cccc}
0 & 0 & 0 & 0\\
0 & \varepsilon B_{N} & -\varepsilon B_{N} & 0\\
0 & -\varepsilon B_{N}  & \varepsilon B_{N} & 0\\
0 & 0 & 0 & \varepsilon^2 C_{N} \\
\end{array}
\right )
\end{equation*}
where $P^{(N)}_{III}=\frac{\varepsilon (2B_{N}+\varepsilon C_{N})}{2^{(N+1)}}$.
Hence the concurrence $C$ is equal to:
\begin{equation*}
C=\frac{2B_{N}}{(2B_{N}+\varepsilon C_{N})}\rightarrow 1\; \; \;\mbox{for}\; \; \varepsilon \rightarrow 0
\end{equation*}
Here we have demonstrated that theoretically the concentration protocol works even if we consider a generic number $N$ of linear interactions between signal and noise, and if only one filtration procedure after all the interactions is applied. It is practically relevant, since it is hard to implement filtration after any particular step of multiple channel. Such result can be experimentally implement at a cost of a low probability of implementation, as shown in Fig.7.

\section{Conclusions and perspectives}
We have reported the experimental demonstration of a concentration
protocol able to restore entanglement on entanglement breaking channel by
measuring the environmental photons. The concentration
procedure can contribute to develop a new class of entanglement sources from noisy devices. Even for the channel preserving entanglement, it can be particularly advantageous to detect environmental state
outgoing from the coupling, because the entanglement
can be increased simply by local filtering on single copy instead of more demanding collective
distillation protocol. The present scheme can be generalized to restore entanglement after consecutive interactions with the different sources of noise. Moreover, a potential application can be envisaged more in quantum entanglement generation in matter devices, where decoherence can be attributed to
coupling of entangled states with other degrees of freedom of the matter. In this framework, a measurement on this available environment can
be adopted to retrieve entanglement. On the fundamental side the present
experiment provide test of important elements to overcome decoherence effects, a subject
of renewed interest in the last year \cite{Alme07} \cite{Konr08}.

R.F. would like to acknowledge AvH Foundation and Grant 202/08/0224 of GACR.

\end{document}